\documentclass[preprint,showpacs,preprintnumbers,amsmath,aps,prd]{revtex4-1} 
\usepackage{graphicx}

\begin{document}
\title{Holographic fermions in asymptotically scaling geometries with hyperscaling violation}
\author{Zhong-Ying Fan}\email{zhyingfan@gmail.com}
\affiliation{Department of Physics, Beijing Normal University, Beijing 100875, China}

\date{\today}

\begin{abstract}
We investigate holographic fermions in general asymptotically scaling geometries with hyperscaling violation exponent $\theta$, which is a natural generalization of fermions in Lifshitz spacetime. We prove that the retarded Green functions in this background satisfy the ARPES (angle-resolved photoemission spectroscopy) sum   rules by introducing a dynamical source on a UV brane for zero density fermionic systems. The big difference from the Lifshitz case is that the mass of probe fermions decoupled from the UV theory and thus has no longer been restricted by unitarity bound. We also study finite density fermions at finite temperature, with dynamical exponent $z=2$. We find that the dispersion relation is linear but the logarithm of the spectral function is not linearly related to the logarithm of $k_\bot =k-k_F$, independent of charge $q$ and $\theta$. Furthermore, we show that with the increasing of charge, new branches of Fermi surfaces emerge and tend to gathering together to form a shell-like structure when the charge reaches some critical value beyond which a wide band pattern appears in the momentum-charge plane. However, all sharp peaks will be smoothed out when $\theta$ increases, no matter how much large the charge is.
\end{abstract}
\maketitle
\section{Introduction}
As is known to all, the strongly coupled field theories can only be studied by additional assumptions, hypothesis and numerical simulations, without  analytical investigations and proof due to the non-perturbative nature. Despite some successes, there however is still lacking of a systematical way and  especially, there emerges new materials in recent years, including the cuprate superconductors and strange metals which seem to lie outside this approach. Fortunately, the AdS/CFT correspondence \cite{1,2,gubser} provides a completely new way to study these strongly interacting theories by constructing the corresponding classical gravity dual which is the only assumption in this theoretical framework. In particular, one can now derive and show the behavior of the Green's functions both analytically and numerically in the boundary theory \cite{3,4,5,6,7,8,9,10,11,12,13,14,15,Reza} by adding probe scalar or spinor fields in the bulk. One of the most studied class of gravity backgrounds is the one with asymptotically anti-de sitter (AdS) spacetime which is dual to a conformal field theory in the UV limit. \\
Recently, people have also constructed the gravity duals for non-relativstic field theories \cite{16,17,18,19,20,21,22} which is clearly necessary and important to investigate the finite density systems in condense-matter theory, thus called AdS/CMT correspondence. The space-time now exhibits anisotropic scale invariance with Lifshitz dynamical scaling exponent $z\neq 1$ ($z=1$ reduces to the AdS case), and thus is known as Lifshitz space-time. More recently, it has been extended to more general scaling geometries with hyperscaling violtion exponent by studying the standard Einstein-Maxwell-dilation theory in the bulk \cite{23,kiritsis,24,25,juan}. The metric behaves like
\begin{equation} ds^2=-\frac{dt^2}{r^{2m}}+r^{2n}dr^2+\frac{dx_i^2}{r^2} \label{1} \end{equation}
where $i=1,2,...,d$ is space index, $m$ and $n$ are related to dynamical exponent $z$ and hyperscaling violation exponent $\theta$ by
\begin{equation}
  z =\frac{m+n+1}{n+2}\ ,\quad   \theta =\frac{n+1}{n+2}\cdot d        \label{2}\end{equation}
Note that $n=-2$ corresponds to a class of spacetime which is conformally related to $AdS_2\times R_2$ with the locally critical limit $z\rightarrow \infty,\   \theta\rightarrow -\infty, \mbox{while}\  z/\theta$ fixed to be a constant\cite{26}. And  when $n=-1$, the metric reduces to the pure Lifshitz spacetime. Furthermore, in order to obtain a stable theory, we will always have \cite{27} $m \geq n+2$, and $m\geq 0$ or equivalently $\theta < d, z\geq 1+\theta/d$ \cite{28}.\\
The metric transforms as
\begin{equation} t \rightarrow \lambda^z t,\ x_i\rightarrow \lambda x_i,\ r\rightarrow \lambda^{\frac{d-\theta}{d}} r,\ ds\rightarrow \lambda ^{\frac{\theta}{d}} ds\label{3} \end{equation}
Hence the background is not scale invariant. And the dual theory in the boundary typically has a non-trivial dimensional parameter below which such behavior emerges. Above this dynamical scale, generally speaking, the metric (\ref{1}) will no longer be a good description probably.

On the other hand, in the standard AdS/CFT correspondence, the probe fermion in the bulk typically corresponds to composite operators whose correlation function obeys modified sum-rules which differs from the ARPES sum-rule\cite{daniel}(see eq.(\ref{sumrule})), obeyed by the single-particle Green's function, as emphasized in Ref.\cite{29,umut}.
\begin{equation} \frac 1\pi \int_{-\infty}^{+\infty}\mathrm{d}\omega \mathrm{Im}[G(\omega,\vec{k})]=1 \label{sumrule}\end{equation}
In order to obtain the single-particle correlation function, Umut $\mathrm{G\ddot{u}}$rsoy et al constructed a modified holographic prescription by introducing a dynamical source on a UV cut-off surface close to the boundary, which actually creates a UV dimensionful scale by hand which does allow for the non-trivial behavior of the Green's function extracted from holography. More precisely, a kinetic term for the bulk field is placed on the UV cut-off surface, instead of the stationary one as is well-known. When integrating over the boundary value of the bulk field via alternative quantization, one finds the two-function of the dynamical source which is proposed to be the holographic dual of the elementary fermion in the boundary. It has been shown that the Green's function obtained in this procedure in asymptotically Lifshitz spacetime does obey the non-trivial sum-rule (\ref{sumrule}) and the Kramers-Kroning relation, with no pole located in the upper half plane of frequency\cite{29}.

In this paper, we will generalize the ARPES sum-rule in asymptotically scaling geometries with hyperscaling violation where we in fact have assumed that the dynamical scale hidden in the metric (\ref{1}) is as the same order as the UV cut-off. We also study holographic fermions at finite density in this background. The rest of this paper is organized as follows: In section 2, we briefly review the charged black hole solutions with the metric (\ref{1}) asymptotically  in standard Einstein-Maxwell-dilaton theory. We introduce two gauge fields one of which behaves finite in the UV limit, thus chemical potential can be well defined. And we study the condition of extracting well fermionic operator dimension in the UV limit. In section 3, we compute the Green's function for probe fermions at zero density and prove the ARPES sum-rule at the zero temperature limit. In section 4, we investigate the properties of holographic fermions at finite density by solving the Dirac equation numerically. Finally, we present a brief conclusion.

\section{Preliminary}
\subsection{Scaling geometries with hyperscaling violation}
We start from the standard Einstein-Maxwell-dilaton theory in $d+2$ dimensional spacetime
\begin{equation} S=\int \mathrm{d}^{d+2}x \sqrt{-g}\ [R-2(\partial{\phi})^2-V(\phi)-\frac{\kappa^2}{2}Z(\phi)F^2-\frac{\kappa^2}{2}H^2]
\label{action}\end{equation}
where the AdS radius is set to $1$. This action has been widely studied with great details in Ref.\cite{23,24,28,kiritsis}. We list the solutions in the following
\begin{equation} ds^2=-r^{-2m}h(r)dt^2+r^{2n}h^{-1}(r)dr^2+\frac{dx^2_i}{r^2}\ ,\qquad h(r)=1-(\frac{r}{r_h})^{\delta} \label{ds2}\end{equation}
where $\delta=m+n+d+1$, $r_h$ is the location of the horizon. And
\begin{equation} F^{rt}=F_0r^{(m-n+d)}Z^{-1}(\phi)\ ,\qquad H^{rt}=H_0r^{(m-n+d)} \label{fh}\end{equation}
where $F_0$, $H_0$ are constants which are proportional to the conserved charges carried by the black brane. And
\begin{equation} \phi=k_0\log{r}\ ,\qquad \quad k_0=\sqrt{\frac d2 (m-n-2)} \label{phi}\end{equation}
\begin{equation} V(\phi)=-V_0 e^{-\beta \phi}\ ,\qquad V_0=\delta(m+d-1)\ ,\qquad  \beta=\frac{2(n+1)}{k_0} \end{equation}
\begin{equation} Z^{-1}(\phi)=Z_0 e^{-\alpha \phi}+Z_1\ ,\ \alpha=\frac{2(n+d+1)}{k_0}\ ,\ Z_0=\frac{\delta(m-1)}{\kappa^2F_0^2}\ ,\ Z_1=-\frac{H_0^2}{F_0^2} \end{equation}
The Hawking temperature of the black brane is give by
\begin{equation} T=\frac{\delta}{4\pi } \frac{1}{r_h^{(m+n+1)}}\label{t}\end{equation}
For a physically stable theory, $k_0$ is real and one finds $m\geq n+2$, which is consistent with the condition that we mentioned in section 1. Note when $r\rightarrow 0$, both the dilaton and the field strength $F^{\mu\nu}$ diverges at the boundary such that chemical potential can't be defined properly. To obtain a well definition for the finite density, we introduce another gauge field H, $H=dB$. where $B_\mu$ is the corresponding gauge potential. From eq.(\ref{fh}), we find
\begin{equation} B(r)=\mu (1-\frac{r^{(d-m+n+1)}}{r_h^{(d-m+n+1)}})\ dt \label{b} \end{equation}
where $\mu$ is the boundary chemical potential. Requiring $H^{rt}$ behaves regularly in the UV limit, we find
\begin{equation} 2\leq m-n \leq d\ ,\qquad d\geq 3 \label{mn}\end{equation}
which leads to a constraint condition on $(z,\theta)$. For instance, when $d=3$, $z\leq 2$, $\theta < 3$.
Notice that when $n=-1$, $m=z$, above solutions reduce to the Lifshitz case.
The diverging asymptotic behavior needs to be properly treated by a holographic renormalization procedure which we won't discuss in this paper. Since the fermions we consider do not couple to the dilaton and the F field, they will be regarded as purely `background' and neglected.
\subsection{Holographic fermions in asymptotically scaling geometries }
In order to investigate holographic fermions in asymptotically scaling geometries, one needs to extract the fermionic operator dimension in the asymptotic limit $r\rightarrow 0$ at least, as in the standard holography. For this purpose, we first consider the fermions at zero density. The fermionic action in the bulk is given by
\begin{equation} S_f[\Psi]=ig_f \int \mathrm{d}^{d+2}x\sqrt{-g}\ \overline{\Psi}(\Gamma^a \mathcal{D}_a-M)\Psi+S_{bdy}[\Psi] \label{faction}\end{equation}
\begin{equation} S_{bdy}[\Psi]=ig_f\int_\epsilon \mathrm{d}^{d+1}x\sqrt{-g_\epsilon}\sqrt{g^{rr}}\overline{\Psi}_+\Psi_-\label{bdy}\end{equation}
where $\overline{\Psi}=\Psi \Gamma^t$, $\mathcal{D}_a=(e_a)^\mu D_\mu$, with $D_\mu=\partial_\mu+\frac 14 \omega_{\mu ab}\Gamma^{ab}$, and $\Gamma^a$ are the $d+2$ dimensional gamma matrices, $\Gamma^{ab}=\frac 12 [\Gamma^a, \Gamma^b]$, $M$ is the mass of the probe fermion in the bulk. $(e_a)^\mu$ are the vielbeins which can be chosen by
\begin{equation} (e_a)^\mu=\sqrt{|g^{\mu\mu}|}(\frac{\partial}{\partial x^\mu})^a \end{equation}
And $\omega$ is the spin connection whose nonzero components are
\begin{equation} \omega_{ttr}=-\omega_{trt}=-\sqrt{g^{rr}}\partial_r\sqrt{-g_{tt}} \nonumber\end{equation}
\begin{equation} \omega_{iir}=-\omega_{iri}=\sqrt{g^{rr}}\partial_r\sqrt{g_{ii}} \end{equation}
Furthermore, $g_\epsilon$ is the determinant of the induced metric on the constant $r$ slice, $r=\epsilon$. $\Psi_\pm$ is defined by
\begin{equation} \Psi_\pm=\frac 12 (1\pm \Gamma^r) \Psi\ ,\qquad \Gamma^r \Psi_\pm=\pm \Psi_\pm  \end{equation}
$S_{bdy}[\Psi]$ is the boundary terms introduced to ensure a well defined variational principle for the total action\cite{30,31}.
The Dirac equation reads
\begin{equation} (\Gamma^a \mathcal{D}_a-M)\Psi=0 \label{dirac} \end{equation}
In the asymptotic limit eq.(1), one readily finds
\begin{equation} [r^{-n}\Gamma^r(\partial_r-\frac{m+d}{2r})+r^m \Gamma^t\partial_t+r\Gamma^{x_i}\partial_{x_i}-M]\Psi(r,\vec{x})=0 \label{dirac2}\end{equation}
Assume that the Dirac field behaves like
\begin{equation} \Psi(r,\vec{x})\rightarrow r^\Delta (\Psi_0(\vec{x})+r\Psi_1(\vec{x})+...),\quad \mbox{when}\ r\rightarrow 0 \label{psi}\end{equation}
where $\Delta$ is the fermionic operator dimension. Substituting eq.(\ref{psi}) into eq.(\ref{dirac2}), one finds that if and only if $n\geq -1$, $\Delta$ can be extracted from asymptotic analysis. When $n=-1$, $\Delta=(z+d)/2\pm M$, which is compatible with the Lifshitz case. When $n> -1$ i.e. $\theta >0$, $\Delta=(m+d)/2$, which is exactly the generalized case we are interested in. The unitarity bound requires $\Delta \geq d/2$ which leads to $m\geq 0$ that is the physically sensible condition we mentioned previously. Hence, the bound is satisfied automatically now. Moreover, the mass of the probe fermion decouples from the UV dimension which will play an important role in the proof of ARPES sum-rule, as we will show in the next section.

Take a Fourier transformation
\begin{equation} \Psi(r,x_\mu)=(-gg^{rr})^{-\frac 14}e^{-i\omega t+ik_ix^i} \psi(r,k_\mu)\ ,\quad k_\mu=(-\omega,\ \vec{k})  \end{equation}
Since the theory is rotational invariant, the momentum can be taken along $x_1$ direction $\vec {k}=k \vec {e}_1$, where $\vec {e}_1$ is the unit vector parallel to the $x_1$ direction. Furthermore, we choose the gamma matrices as follows\cite{3,31}
\begin{equation} \Gamma^r=
\left( \begin{array} {ccc}
   -\sigma^3 & 0\\
   0 & -\sigma^3
\end{array}\right)
\ ,\quad
\Gamma^t=
\left( \begin{array}{ccc}
  i\sigma^1 & 0 \\
  0 & i\sigma^1
\end{array}\right)
\ ,\quad
\Gamma^{x_1}=
\left( \begin{array}{ccc}
    -\sigma^2 & 0 \\
    0 & \sigma^2
\end{array}\right)
\end{equation}
where $\sigma$ are Pauli matrices. And we set
\begin{equation}
\psi= \left(\begin{array}{ccc} \psi_+ \\ \psi_-\end{array}\right)\ ,\qquad \psi_\pm=\left(\begin{array}{ccc}u_\pm \\ d_\pm \end{array}\right)
\end{equation}
The Dirac equation yields
\begin{equation}\label{xi}
\sqrt{g^{rr}}\partial_r \xi_\pm +2M \xi_\pm=(\omega \sqrt{-g^{tt}} \pm k\sqrt{g^{x_1x_1}})\xi^2_\pm+(\omega \sqrt{-g^{tt}} \mp k\sqrt{g^{x_1x_1}})
\end{equation}
where $\psi_-(r,k_\mu)=-i\xi(r,k_\mu)\psi_+(r,k_\mu)$ and $\xi_+=iu_-/u_+,\ \xi_-=id_-/d_+$. $\xi_\pm$ are the eigenvalues of the matrices $\xi$. In the standard AdS/CFT correspondence, when $r\rightarrow 0$, $\psi_-(r,k_\mu)\sim r^{\Delta_-}A(k_\mu)$, $\psi_+(r,k_\mu)\sim r^{\Delta_+}B(k_\mu)$ (The other mode is analyzed in Appendix) whereas the Green function is given by $G_{O_-}(k_\mu)=i \frac{A(k_\mu)}{B(k_\mu)}$ which will lead to $G_{O_-}=\lim_{r\rightarrow 0}r^{(\Delta_+-\Delta_-)}\xi(r,k_\mu)$. In our background, $\Delta_+=\Delta_-=\frac{d+m}{2}$, thus we readily obtain the retarded correlation functions of fermionic operators by imposing in-falling boundary conditions for $\psi$ at the event horizon\cite{26}
\begin{equation} G_{O_-}(\omega,\vec{k})=\lim_{r\rightarrow 0} \xi(r,\omega,\vec{k})\ ,\qquad \xi(r_h,\omega,\vec{k})=i\ ,\quad\mbox{for}\  \omega\neq 0 \label{gr}\end{equation}
where $O_-$ is the fermionic operator dual to $\psi_+$. Note that eq.(\ref{gr}) is always true for arbitrary fermion mass $M$ in contrast to the Lifshitz cases where a factor $r^{-2M}$ appears in the righthand side of the limit to obtain a finite result \cite{38}. For finite density systems, eq.(\ref{xi}) is still valid, with $k_\mu$ replaced by $k_\mu-q B_\mu$, where $q$ is the charge carried by the probe fermions under the gauge field $B_\mu$. Furthermore, eq.(\ref{xi}) leads to
\begin{equation}\xi_\pm(r,\omega,k,M)=\xi_\pm(rk^{\frac{1}{n+2}},\frac{\omega}{k^z},\frac{M}{k^{\theta/d}})
=\xi_\pm(r\omega^{\frac{1}{(n+2)z}},\frac{k}{\omega^{\frac 1z}},\frac{M}{\omega^{\theta/(dz)}}) \label{xixi} \end{equation}
when $n=-1$, $\theta=0$, eq.(\ref{xixi}) reduces to the Lifshitz case.
\section{ARPES sum-rules}
From eq.(\ref{xi}) and eq.(\ref{gr}), one can easily drive the two point function of the fermionic opertor $O_-$ in the boundary. However, as we reviewed in the previous section, the Green's function derived in this standard holographic procedure doesn't obey the ARPES sum-rules eq.(\ref{sumrule}). It was first proposed in Ref.\cite{29} that a modified approach which introduces a dynamical source on a UV cut-off surface does allow for the extraction of the single-particle Green's function that satisfies eq.(\ref{sumrule}). The action of the source term is\cite{29}
\begin{equation} S_{UV}[\Psi_+]=Z\int_{r=\epsilon}\mathrm{d}^{d+1}x \sqrt{-g_\epsilon}\ \overline{\Psi}_+(i\Gamma^a (e_a) ^\mu \partial_\mu)\Psi_+ \label{source}\end{equation}
Hence the total action \cite{39} is
\begin{equation} S_{tot}[\Psi]=S_f[\Psi]+S_{UV}[\Psi_+] \label{totalaction}\end{equation}
We focus on $d=3$ case. The corresponding retarded Green's function extracted from this action is as follows
\begin{equation} G_R(\epsilon, k_\mu)=-(\omega-\sqrt{-g_{tt}g^{x_1x_1}}(\vec{\sigma}\cdot \vec{k})+\frac{g_f}{Z}\sqrt{-g_{tt}g^{rr}}\xi(\epsilon, k_\mu))^{-1} \end{equation}
where we have redefined the Dirac field as $\Psi_+\rightarrow Z^{-\frac 12}(g^{tt}g_\epsilon)^{-\frac 14}\Psi_+$. In the zero temperature limit, we obtain
\begin{equation} G_R(\epsilon, k_\mu)=-(\omega-\epsilon^{1-m}(\vec{\sigma}\cdot \vec{k})+\frac{g_f}{Z}\epsilon^{-(n+m)} \xi(\epsilon, k_\mu))^{-1}  \label{zerogr}\end{equation}
In the end, we need to take the limit $\epsilon\rightarrow 0$ to romove the UV cut-off. Obviously, the last two terms in eq.(\ref{zerogr}) are divergent in this limit. The divergence in the third term can be absorbed in the constant $g_f$ by a redefinition $\tilde{g}_f=g_f\epsilon^{-(n+m)}/Z$. In order to keep $\tilde{g}_f$ finite to arbitrary order, we work in a double scaling limit $\epsilon\rightarrow 0,\ g_f\rightarrow 0,\ g_f\epsilon^{-(n+m)}=const$. Since the theory is not scale invariant (see eq.(\ref{3})), the second term needs to be treated carefully. Fortunately, the scaling relation $\omega \propto k^z$ still holds and is preserved under the scale transformation eq.(\ref{3}). Hence, the kinetic action for a single component $\psi$ of a fermion also reads $S_{kin}\sim \int \mathrm{d}\omega\ \mathrm{d}^d k\ \psi^*(\omega+\eta k^z)\psi$. When the UV cut-off of the spatial momentum $\Lambda_k \sim \epsilon$ is taken, the Lifshitz scaling is broken and the relevant terms contribution which looks like $S_{rel}\sim \int \mathrm{d}\omega\ \mathrm{d}^d k\ \psi^*(\tilde{\eta}k)\psi$ needs to be considered in such a way that it will maintain this scaling in the IR region. Thus, we obtain
\begin{equation} G_R(\omega,\vec{k})=-(\omega+\eta \vec{\sigma}\cdot \vec{k} k^{z-1}+\tilde{g}_f\xi(\omega,\vec{k}))^{-1}\label{finalgr} \end{equation}
where $\eta$ is a constant, $\xi(\omega,\vec{k})=\lim_{\epsilon \rightarrow 0}\xi(\epsilon,\omega,\vec{k})$.

In order to prove the sum-rule eq.(\ref{sumrule}), we first consider massless fermions with zero momentum i.e. $M=0$, $k=0$. From the Dirac equation and in-falling condition, we obtain\cite{3}
\begin{equation} \xi(\omega, \vec{0})=i \end{equation}
Hence, the Green's function of the massless fermions with zero momentum is
\begin{equation} G_R(\omega,\vec{0})=-\frac{1}{\omega+i\tilde{g}_f} \label{gr0}\end{equation}
which clearly leads to
\begin{equation} G_R^\dag(\omega,\vec{0})=-G_R(-\omega,\vec{0}) \end{equation}
Moreover, from eq.(\ref{finalgr}), we find
\begin{equation} \mbox{Tr}G_R(\omega,\vec{k})=-\frac{2\omega+\tilde{g}_f(\xi_+(\omega,\vec{k})+\xi_-(\omega,\vec{k}))}
{(\omega+\tilde{g}_f\xi_+(\omega,\vec{k}))(\omega+\tilde{g}_f\xi_-(\omega,\vec{k}))-\eta^2k^{2z}} \end{equation}
On the other hand, from eq.(\ref{xi}) $\xi(\omega,\vec{k})$ satisfies the following properties\cite{3}
\begin{equation} \xi^\dag(\omega,\vec{k})=-\xi(-\omega,-\vec{k})\ ,\quad \xi_\pm(\omega,\vec{k})=\xi_\mp(\omega,-\vec{k}) \end{equation}
Thus, one can verify that in general we have
\begin{equation} \mbox{Tr}G^\dag_R(\omega,\vec{k})=-\mbox{Tr}G_R(-\omega,\vec{k}) \end{equation}
From eq.(\ref{gr0}), there exists only one pole in the negative imaginary axis provided that $\tilde{g}_f$ is positive definite. Clearly it satisfies the Kramers-Kronig relations. Now we readily obtain
\begin{equation} \frac{1}{2\pi }\int_{-\infty}^{+\infty}\mathrm{d}\omega\; \mbox{Im\;Tr}G_R(\omega,\vec{0})=1 \label{sum}\end{equation}
In the general case $M\neq 0 \neq k$, we have no way to drive the analytic structure of $\xi(\omega,\vec{k})$ and have to assume that all non-analyticity of eq.(\ref{finalgr}) located in the lower half-plane, which is physically sensible for a theory with well causality.
In this case, to perform the calculation of the expression (\ref{sum}), we can use an infinite semi-circle, denoted as $C$ in the upper half-plane to connect the both ends of the real axis to form a closed curve. On this semi-circle, $\omega\rightarrow \infty$ with $k,\ M$ finite. On the other hand, from eq.(\ref{xixi}), we find
\begin{equation} \xi(\omega,k,M)=\xi(\frac{k}{\omega^{1/z}},\frac{M}{\omega^{\theta/(dz)}}) \label{40}\end{equation}
Note that $n> -1$ leads to $\theta > 0$. Clearly, this expression determines $\xi(\omega,k,M)$ that the limit $\omega\rightarrow \infty$ with $k,\ M$ finite is equivalent to the limit $k\rightarrow 0,\ M\rightarrow 0$ with $\omega$ finite. Hence, on the infinite semi-circle, eq.(\ref{finalgr}) will reduce to eq.(\ref{gr0}). Thus, we obtain
\begin{equation} \frac{1}{2\pi }\int_{-\infty}^{+\infty}\mathrm{d}\omega\; \mbox{Im\;Tr}G_R(\omega,\vec{k})=
                 \frac{1}{2\pi }\int_C\mathrm{d}\omega\; \mbox{Im\;Tr}G_R(\omega,\vec{k})=1
\label{finalsumrule}\end{equation}
This is the sum-rule eq.(\ref{sumrule}) obeyed by the single-particle Green's function. Thus, by introducing a dynamical source on a UV cut-off surface, we indeed extract the modified holography which is valid for the elementary fields, instead of the composite operators in the standard AdS/CFT correspondence, as we expected at the very start. Notice that from eq.(\ref{40}) the mass of the probe fermions decouples in the large frequency limit which leads to the fact that the sum-rule holds for arbitrary fermion mass, without generating any instability. This is qualitatively different from the Lifshitz case\cite{29} where the mass is restricted in an interval $(-z/2,z/2)$ to meet the unitarity bound which is however automatically satisfied and has no power in our asymptotically scaling geometries with hyperscaling violation. It isn't surprising since the mass decouples from the UV dimension and does not appear in the asymptotic expansion of the fermion field, as it was shown eq.(\ref{psi}). We point out that this nontrivial behavior of the fermion mass is also true for scalar and spinor fields in our background. We don't know the physical origin in a deeper level. One of possible explanations is that the dual theory in the boundary is not scale invariant from the UV fixed point. The dynamical scale below which the hyperscaling violation emerges is the same order of the UV cut-off, which probably results to the decouple of the bulk mass in the operator dimension and instead the hyperscaling violation appears in the dimension.
\section{Finite density systems}
We now return to the holographic fermions at finite density in the background (\ref{ds2}). The charged black brane has zero entropy density at zero temperature limit and doesn't admit extremal solutions. Despite this point, we will study the fermions at low temperature and explore the existence of Fermi surfaces. Since the system has finite density, we need to replace $\omega$ by $\omega-qB_t$ in eq.(\ref{xi}). Furthermore, the Green's function $G_{O_-}(\omega,k)$ is a two by two matrice whose eigenvalue is $G_{11}$ and $G_{22}$ which are related by $G_{11}(\omega,k)=G_{22}(\omega,-k)$. We will focus on $G_{22}$ and drop the subscript in the following.

\begin{figure}[tbp]
\includegraphics[width=7.5cm]{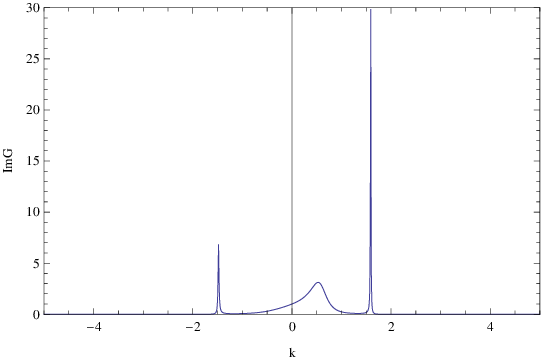}
\includegraphics[width=7.5cm]{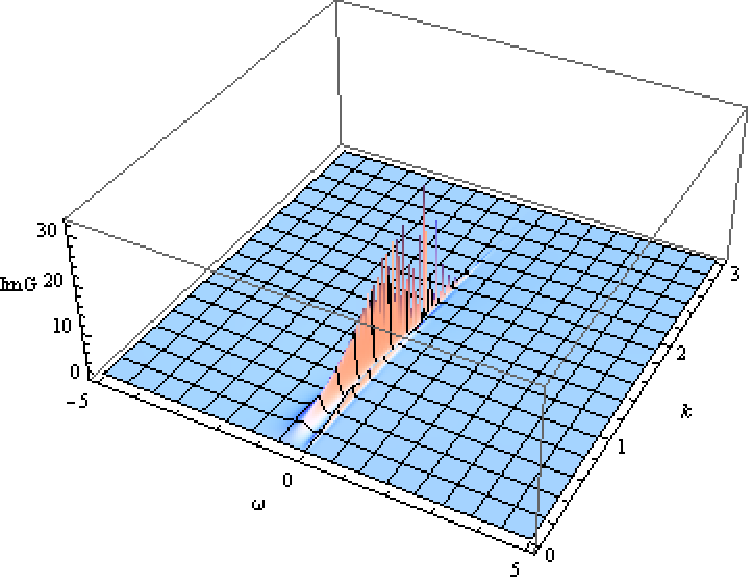}
\caption{The plot of $\mbox{Im}G$, for $q=2.5,\ (\omega=1\times 10^{-12})$. }
\end{figure}
\begin{figure}[tbp]
\includegraphics[width=7.5cm]{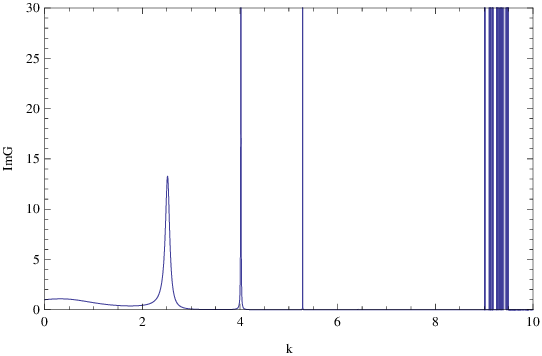}
\includegraphics[width=7.5cm]{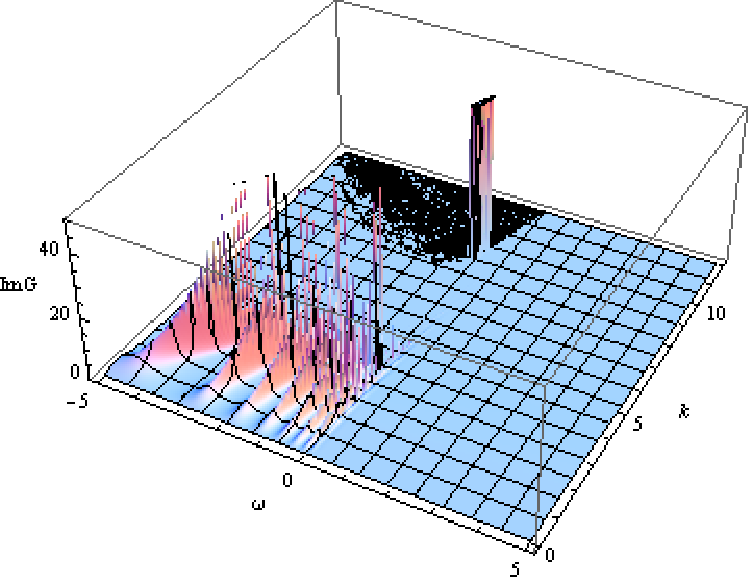}
\caption{The plot and 3D plot of $\mbox{Im}G$ for $q=12$ $(\omega=1\times 10^{-12})$. }
\end{figure}
For convenience to perform numerical calculation, we set $d=3,\ M=0,\ z=2,\ \mu=1,\ T=\frac{1}{16\pi}$. The remained free parameters are $n$ and $q$. Especially, the hyperscaling violation exponent is expressed as $\theta=3(n+1)/(n+2)$, varying with $n$ only. Therefore, we won't mention $\theta$ again, unless it is necessary to.

We first focus on $n=0$ case. The properties of the spectral function $\mbox{Im}G$ are summarized as follows:
\begin{figure}[tbp]
\includegraphics[width=7.5cm]{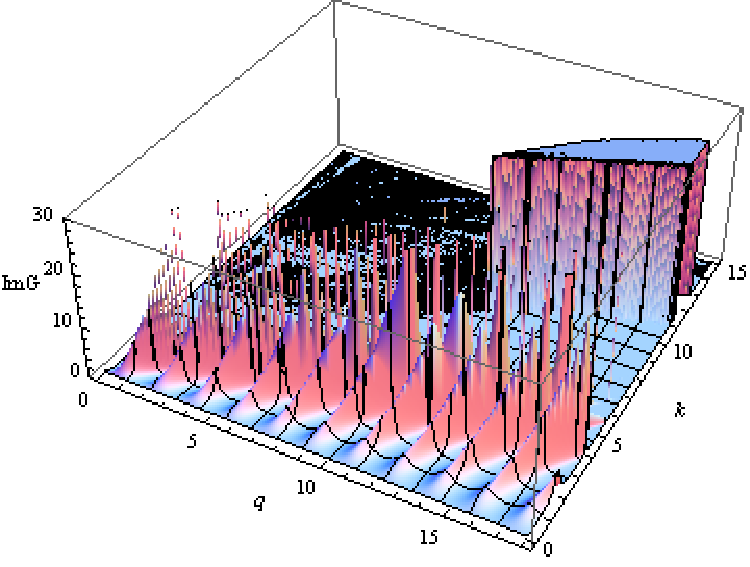}
\includegraphics[width=7.5cm]{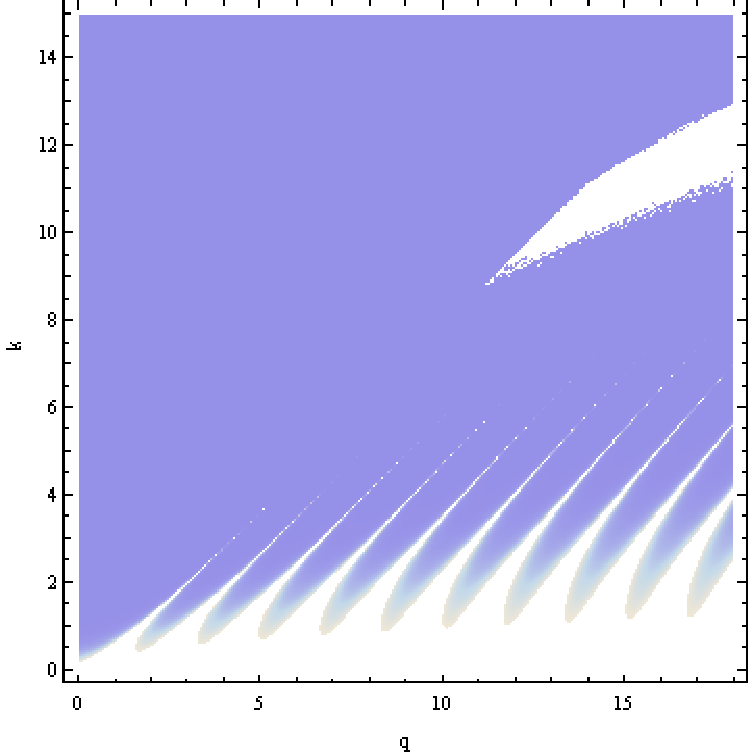}
\caption{The 3D and density plots of $\mbox{Im}G$ for $n=0$ $(\omega=1\times 10^{-12})$.}
\end{figure}
\paragraph{}
It is shown that a sharp quasiparticle like peak appears at some momentum in $\omega\rightarrow 0$ limit in Figure.1. The peak is almost a delta function, implying the existence of a Fermi surface. The corresponding Fermi momentum is $k_F=1.58848138$ for $q=2.5$. We also obtain $k_F=1.20444080$ for $q=2$ and $k_F=1.98251000$ for $q=3$. Hence, the Fermi momentum increases linearly with the charge $q$, compatible with the results obtained in charged black holes in Lifshitz and AdS spacetime\cite{3,12}.
\paragraph{}
 From Figure.2 and Figure.3 \cite{40}, we find that when the charge is sufficiently large, there even emerges Fermi-shell like structure in the spectral function which contains a large number of sharp peaks with tiny steps in some narrow interval of momentum. The spectral function behaves highly singular around the shell such that on one hand, it takes an absolute dominant position instead of the sharp peaks in the lower charge, with a ratio of order $10^{45}$ in our numerics; on the other hand, the shell appears discontinuously and suddenly during the increasing of the charge, as shown in Figure.3. Clearly, there exists a critical charge to allow for the emergent of the `q-band' pattern, $q_c\approx 11.2\ \mbox{for}\ n=0$. The band goes like a line, indicating that the Fermi momentum is linearly related to the charge,  enjoying the same behavior of the lower charge peaks. This may not surprising, since the shell is made up of plenty of densely gathering sharp peaks, with no qualitative difference from the later case.
\paragraph{}
 In Figure.4, we find that the dispersion relation between $\omega_*(k_\perp)$ and $k_\perp=k-k_F$ is linear, independent of the charge $q$, where $\omega_*(k_\perp)$ is the location of the maximum of the spectral weight $\mbox{Im}G$, indicating that the dual liquid is like a Fermi liquid.

\paragraph{}
In Figure.5, we plot the scaling behaviors of the spectral weight at the maximum. The logarithm of $\mbox{Im}G$ is however not linearly dependent on $\log|k_\perp|$, behaving qualitatively different from the published results previously \cite{3,4,5,10,11,12}, where $\mbox{Im}G(\omega_*(k_\perp),k_\perp)\sim k^{-\beta}_\perp$ and $\beta=1$ for the standard Landau-Fermi liquid. In our case, $\beta \rightarrow 0$ with $k_\perp\rightarrow 0$ limit. In this regard, the dual liquid behaves not exactly as a Landau Fermi type. This phenomenon is also observed in charged dilatonic black holes with non-relativistic fermionic fixed point\cite{13}.
\paragraph{}
 The above results stand for the general properties of the systems with finite $n$ ($n> -1)$. However, when $n$ increasing, the sharp peaks will be smoothed out for any fixed charge, as shown in Figure.6. Thus in the large $n$ limit, the hyperscaling violation will play a dominant role instead of the finite density such that there won't exist any Fermi surface. Indeed, $\theta < d$  is a compulsory constraint to allow for a stable theory (see eq.(\ref{2})). Hence, $\theta \rightarrow d$ limit might be a critical point, through which a complete new phase with no Fermi surfaces emerges, via some unkonwn quantum phase transitions.
\begin{figure}[tbp]
\includegraphics[width=5.2cm]{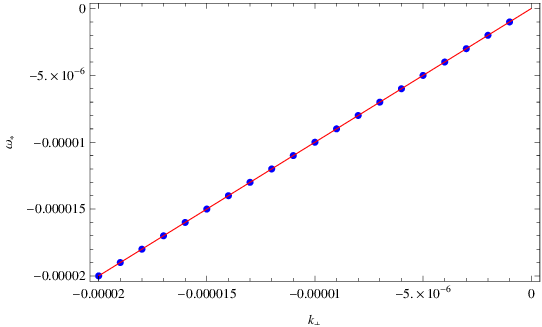}
\includegraphics[width=5.2cm]{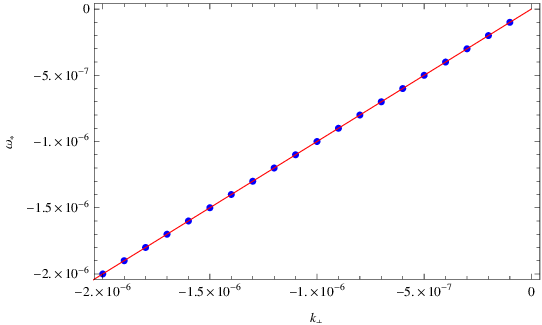}
\includegraphics[width=5.2cm]{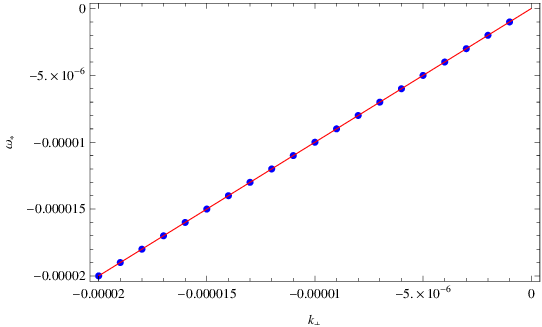}
\caption{The dispersion relation between $\omega_*$ and $k_\perp$, from left to right $q=2\ ,\ 2.5\ ,\ 3$. }
\end{figure}
\begin{figure}[tbp]
\includegraphics[width=5.2cm]{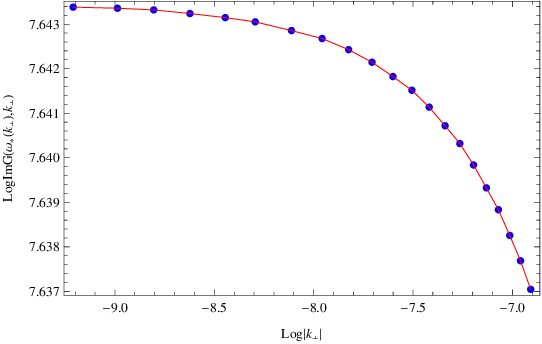}
\includegraphics[width=5.2cm]{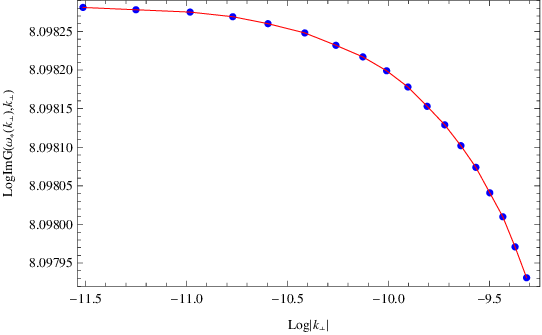}
\includegraphics[width=5.2cm]{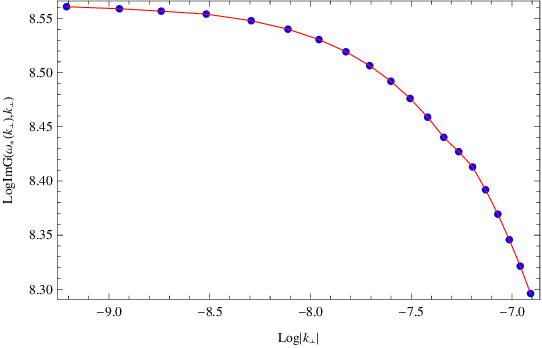}
\caption{The scaling behaviors of the maximum spectral height $\mbox{Im}G(\omega_*(k_\perp),k_\perp)$, from left to right $q=2\ ,\ 2.5\ ,\ 3$. }
\end{figure}
\begin{figure}[htbp]
 \includegraphics[width=7.5cm]{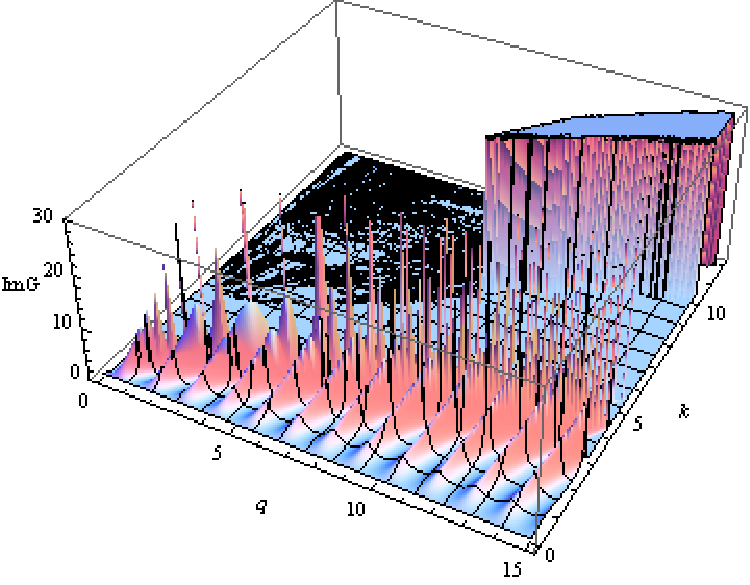}
 \includegraphics[width=7.5cm]{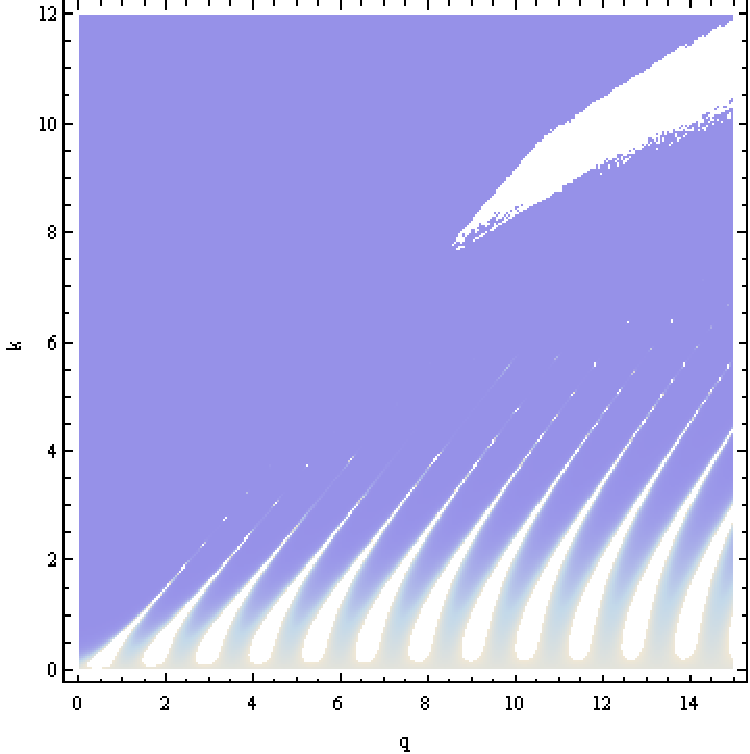}
 \includegraphics[width=7.5cm]{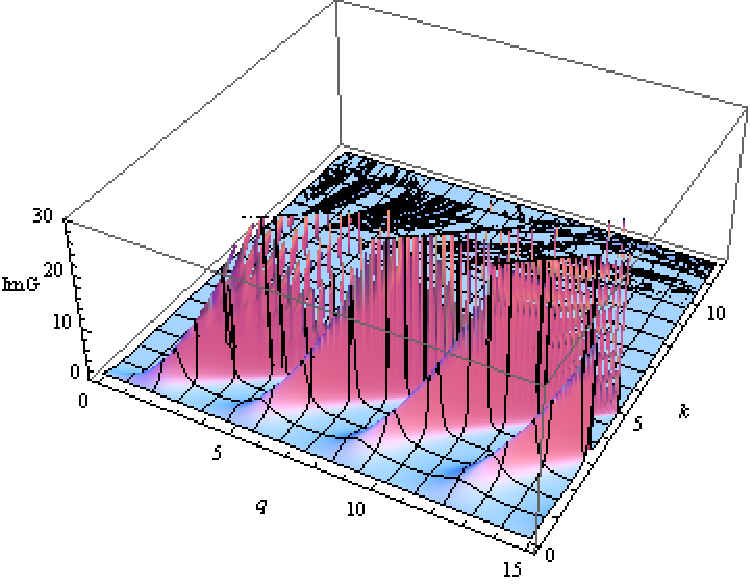}
 \includegraphics[width=7.5cm]{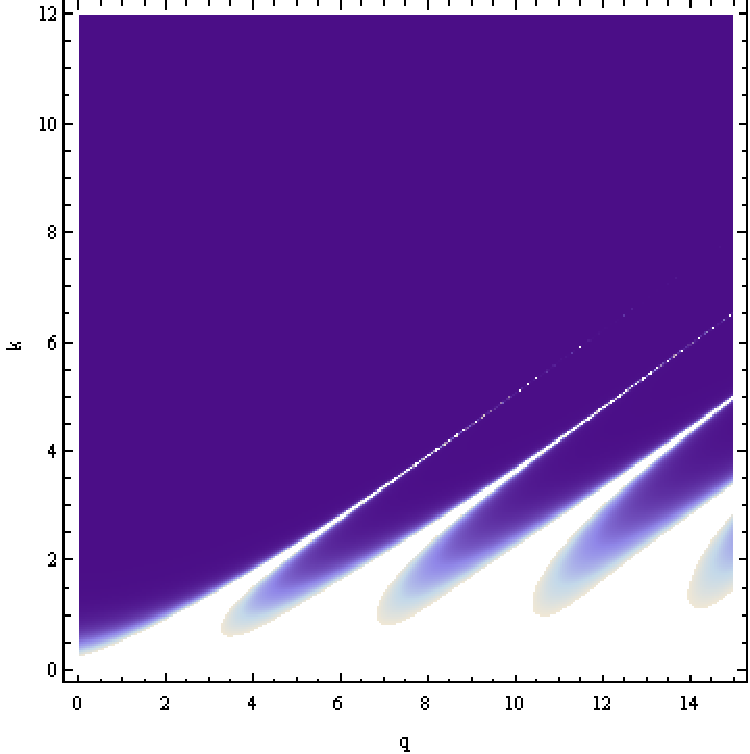}
 \includegraphics[width=7.5cm]{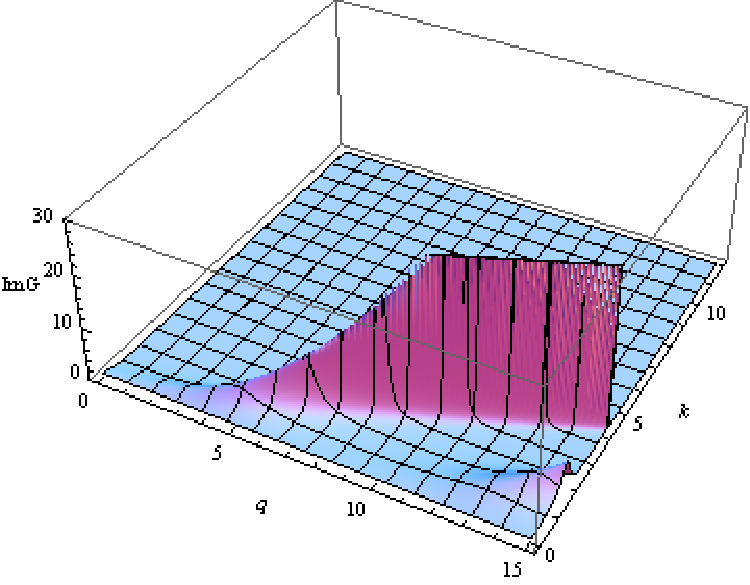}
 \includegraphics[width=7.5cm]{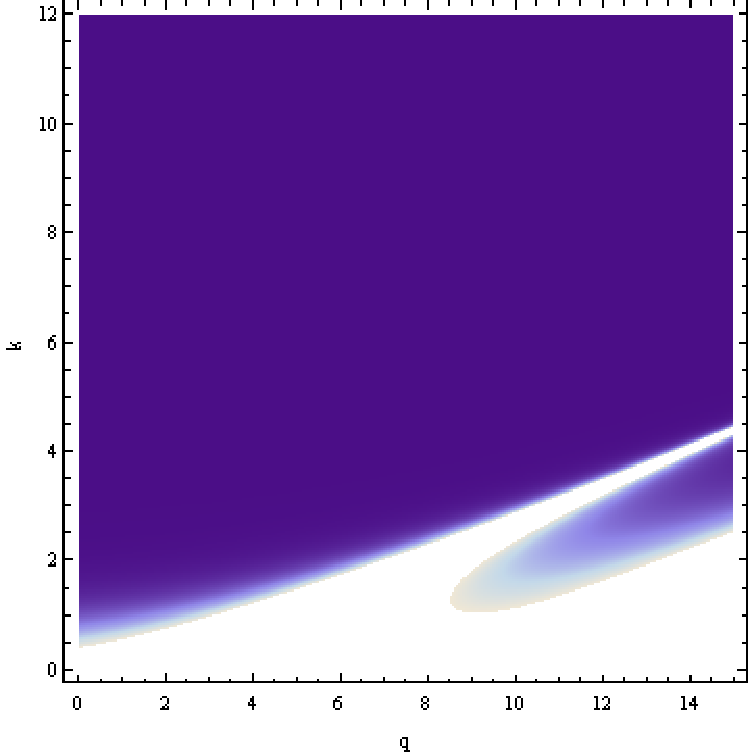}
 \caption{The 3D and density plots of $\mbox{Im}G$, from top to bottom $n=-0.5,\ n=2,\ n=8$.}
\end{figure}
\section{Conclusions}
We have studied holographic fermions in asymptotically scaling geometries with hyperscaling violation $\theta$. In order to extract the fermionic operator dimension for a stable theory by asymptotical analysis, as in the standard holography, we need $0< \theta < d$. We first prove that the Green's function obtained in the modified holography\cite{29} which introduces a dynamical source on a UV cut-off surface close to the boundary does exactly obey ARPES sum-rule in this spacetime. The fermion's mass decouples from the UV sector and no longer imposes restrictions on the sum-rule. \\
On the other hand, we also studied the main features of the fermions at finite density in low temperature. The dual liquid may be a Fermi liquid since the dispersion relation is linear, independent of charge $q$ and hyperscaling $\theta$. However, the maximum height of the spectral function behaves qualitatively different from the usual form $\mbox{Im}G\sim k_\perp^{-\beta}$. In our case, $\beta$ is not a constant and varies with $k_\perp$, especially $\beta\rightarrow 0$ in $k_\perp\rightarrow 0 $ limit. Another peculiar property of our system is the sharp peaks emerging with the increasing of charge will gather together to shape a Fermi-shell like structure. It only happens beyond a critical charge value where a novel band pattern appears on the $k$-$q$ plane. In the end, we also find that the sharp peaks become smooth when $\theta$ increases, indicating that there may exist a new phase which has no Fermi surface in $\theta\rightarrow d$ limit.


\section{Appendix}

The Dirac equation can be reexpressed as follows:

\begin{equation}(\sqrt{g^{rr}}\partial_r-M)\psi_+(r,x)+D_-\psi_-(r,x)=0\  \end{equation}
 \begin{equation}(\sqrt{g^{rr}}\partial_r+M)\psi_-(r,x)+D_+\psi_+(r,x)=0\    \end{equation}

where $\Psi=\left( \begin{array} {c}\psi_+ \\ \psi_- \end{array}\right)$,  $D_\pm=\pm i(\sqrt{-g^{tt}}\gamma^t \omega-\sqrt{g^{xx}}\vec{\gamma}\cdot\vec{k} )$. Here the Gamma matrices are chosen as
\begin{equation} \Gamma^\mu=\gamma^\mu\ ,\qquad \Gamma^r=\gamma^{d+2},\quad \mathrm{when\ d\ odd}\end{equation}
\begin{equation} \Gamma^\mu=\left(\begin{array}{cc}0&\gamma^\mu\\\gamma^\mu&0\end{array}\right) ,\qquad \Gamma^r=\left(\begin{array}{cc}1&0\\0&-1\end{array}\right),\quad \mathrm{when\ d\ even}\end{equation}

where $\gamma^\mu$ denotes the boundary gamma matrices. From these equations, we obtain
 \begin{equation}\psi_+(r,k)\rightarrow r^{\Delta_+} A_+(k)+r^{\Delta_--(n+2)}B_+(k)   \end{equation}
\begin{equation}\psi_-(r,k)\rightarrow r^{\Delta_-} A_-(k)+r^{\Delta_+-(n+2)}B_-(k)    \end{equation}

wehre $A_\pm(k)$ are related to $B_\pm(k)$ from above equations whose explicit form depends on the value of parameters $n$ and $m$. For our case, $n>-1$, $\Delta_\pm=\frac{d+m}{2}$,
\begin{equation} \Delta_+ A_+(k)+i\vec{\gamma}\cdot \vec{k}B_-(k)=0\ ,\quad \Delta_- A_-(k)-i\vec{\gamma}\cdot \vec{k}B_+(k)=0   \end{equation}

when $n=-1$, $m=z>1$, the coefficients of $A_\pm$ should be replaced by $\Delta_\pm\rightarrow \Delta_\pm\mp M$ with $\Delta_\pm=\frac{d+z}{2}\mp M$. Moreover, if $m=z=1$ which is the pure AdS case, one should also replace $\vec{\gamma}\cdot \vec{k}$ by $\gamma\cdot k=-\gamma^t\omega+\vec{\gamma}\cdot \vec{k}$.

\section{Acknowledgments}
I thank Dr.Li Qing Fang and Dr.Wei-Jia Li for sharing their Mathematical codes. I also thank Professor Sije Gao for his encouragement. This work is supported by NSFC Grants NO.10975016 and NO.11235003.


\begin{thebibliography}{99}
\bibitem{1}J.Maldcena, Int.\ J. Theor.\ Phys.\ 38 (1999) 1113 [arXiv:hep-th/9711200].
\bibitem{2}E.Witten, arXiv:hep-th/9802150.
\bibitem{gubser}S.S.Gubser, I.R.Klebanov and A.M.Polyakov, Phys.\ Lett.\ B 428:105-114, 1998 [arXiv:hep-th/9802109].
\bibitem{3}T.Faulkner, H.Liu, J.McGreevy, and D.Vegh, Phys.\ Rev.\ D 83,125002 (2011) [arXiv:0907.2694v2 [hep-th]].
\bibitem{4}N.Iqbal, H.Liu, and M.Mezei, arXiv:1110.3814v1 [hep-th].
\bibitem{5}Hong Liu, John McGreevy and David Vegh, Phys.\ Rev.\ D 83, 065029 (2011) [arxiv:0903.2477 [hep-th]].
\bibitem{6}S.A.Hartnoll and Alireza Tavanfar, Phys.\ Rev.\ D 83,046003 (2011) [arXiv:1008.2828 [hep-th]].
\bibitem{7}S.A.Hartnoll, D.M.Hofman, D.Vegh, JHEP 1108, 096 (2011) [arXiv:1105.3197[hep-th]].
\bibitem{8}H.L\"{u}, Zhao-Long Wang, arXiv:1210.4560 [hep-th].
\bibitem{9}Jun Li, Hai-Shan Liu, H.L\"{u} and Zhao-Long Wang, JHEP 02 (2013) 109 [arXiv:1210.5000 [hep-th]].
\bibitem{10}Jian-Pin Wu, Phys.\ Rev.\ D 84, 064008 (2011) [arXiv:1108.6134 [hep-th]].
\bibitem{11}Jian-Pin Wu, JHEP 07 (2011) 106 [arXiv:1103.3982 [hep-th]].
\bibitem{12}Li Qing Fang, Xian-Hui Ge, Xiao-Mei Kuang, Phys.\ Rev.\ D 86 105037 (2012)[arXiv:1201.3832 [hep-th]].
\bibitem{13}Wei-Jia Li, Ren\'{e} Meyer, Hongbao Zhang, JHEP 01 (2012) 153 [arXiv:1111.3783 [hep-th]].
\bibitem{14}Wei-Jia Li, Hongbao Zhang, JHEP 11 (2011) 018 [arXiv:1110.4559 [hep-th]].
\bibitem{15}Wei-Jia Li, Jian-Pin Wu, Nuclear Physics B 867 (2013) 810 [arXiv:1203.0674 [hep-th]].
\bibitem{Reza}M.Alishahiha, M.Reza, M.Mozaffar, A.Mollabashi, Phys.\ Rev.\ D 86, 026002 [arXiv:1201.1764 [hep-th]].
\bibitem{16}S.Kachru, X.Liu, M.Mulligan, Phys.\ Rev.\ D 78, 106005 (2008) [arXiv:0808.1725 [hep-th]].
\bibitem{17}M.Taylor, arXiv:0812.0530 [hep-th].
\bibitem{18}A.Volovich, C.Wen, JHEP 05 (2009) 087 [arXiv:0903.2455].
\bibitem{19}Robert G. Leigh, Nam Nguyen Hoang, JHEP 11 (2009) 010 [arXiv:0904.4270].
\bibitem{20}M.H.Dehghani, R.B.Mann, R.Pourhasan, Phys.\ Rev.\ D 84, 046002 (2011) [arXiv:1102.0578 [hep-th]].
\bibitem{21}J.Tarr\'{l}o, S.Vandoren, JHEP 09 (2011) 017 [arXiv:1105.6335 [hep-th]].
\bibitem{22}Da-Wei Pang, JHEP 01 (2010) 116 [arXiv:0911.2777].
\bibitem{23}Christos Charmousis, Blaise Gouteraux, Bom Soo Kim, Elias Kiritsis, Rene Meyer, JHEP 11 (2010) 151 [arXiv:1005.4690 [hep-th]].
\bibitem{kiritsis}B.Gout\'{e}raux, E.Kiritsis, JHEP 12 (2011) 036 [arXiv:1107.2116 [hep-th]].
\bibitem{24}Norihiro lizuka, Nilay Kundu, Prithvi Narayan and Sandip P.Trivedi, JHEP 1201 (2012) 094 [arXiv:1105.1162 [hep-th]].
\bibitem{25}Xi Dong, S.Harrison, S.Kachru, G.Torroba and H.Wang, JHEP 06 (2012) 041 [arXiv:1201.1905v4 [hep-th]].
\bibitem{juan}M.Edalati, Juan F.Pedraza, W.T.Garcia, Phys.\ Rev.\ D 87, 046001 (2013) [arXiv:1210.6993 [hep-th]].
\bibitem{26}S.A.Hartnoll and E.Shaghoulian, JHEP 07 (2012) 078 [arXiv:1203.4236[hep-th]].
\bibitem{27}Noriaki Ogawa, Tadashi Takayanagi and Tomonori Ugajin, JHEP 01 (2012) 125 [arXiv:1111.1023v4 [hep-th]].
\bibitem{28}Liza.\ Huijse, S.Sachdev,\ B.Swinger, Phys.\ Rev.\ B 85, 035121 (2012) [arXiv:1112.0573[cond-mat.str-el]].
\bibitem{daniel}D.R.Gulotta, C.P.Herzog, M.Kaminski, JHEP 01 (2011) 018 [arXiv:1010.4806[hep-th]].
\bibitem{29}U.G\"{u}rsoy, E.Plauschinn, H.Stoof, S.Vandoren, JHEP 05 (2012) 018  [arXiv:1112.5074 [hep-th]].
\bibitem{umut}U.G\"{u}rsoy, V.Jacobs, E.Plauschinn, H.Stoof, S.Vandoren, arXiv:1209.2593[hep-th].
\bibitem{30}M.Henneaux,  arXiv:hep-th/9902137.
\bibitem{31}N.Iqbal, H.Liu, Fortschr.\ Phys.\ 57, NO.\ 5-7 (2009) [arXiv:0903.2569 [hep-th]].
\bibitem{38}In Lifshitz theory, $\Delta_\pm=\frac{d+z}{2}\mp M$ which results to $G_{O_-}=\lim_{r\rightarrow 0}r^{-2M}\xi(r,k_\mu)$.
\bibitem{39}Here the summer over the indices are over d+2 dimensions, including the radial direction. But the latter contributions vanishes due to bulk chirality $\overline{\Psi}_+\Gamma^r \Psi_+$.
\bibitem{40}The black parts of the 3D plots in these two figures and Figure.6 are purely numerical noise. We haven't found a proper way to remove the noise but we would like to keep these 3D plots as a comparison with the elegant density plots.

\end{thebibliography}
\end{document}